\documentclass[doublecol]{epl2} 

\usepackage{amsfonts}
\usepackage{amssymb}
\usepackage{bm, bbm, dsfont}
\usepackage{subfigure}
\PassOptionsToPackage{caption=false}{subfig}
\usepackage[T1]{fontenc}
\usepackage[latin2]{inputenc}
\usepackage{amsmath}
\usepackage{latexsym}

\usepackage{color}

\newcommand{\ket}[1]{| #1 \rangle}
\newcommand{\bra}[1]{\langle #1 |}

\def\be{\begin{equation}}
\def\ee{\end{equation}}
\def\ben{\begin{eqnarray}}
\def\een{\end{eqnarray}}
 \def\beqa{\begin{eqnarray}}
\def\eeqa{\end{eqnarray}}
\def\eea{\end{array}}
\def\bea{\begin{array}}
\newcommand{\bei}{\begin{itemize}}
\newcommand{\eei}{\end{itemize}}
\newcommand{\bee}{\begin{enumerate}}
\newcommand{\eee}{\end{enumerate}}

\title{Dynamical Objectivity in Quantum Brownian Motion}
\shorttitle{Title} 

\author{J.~Tuziemski \inst{1,3} \and J.~K.~Korbicz \inst{2,3,1} }
\shortauthor{J.~Tuziemski \etal}

\institute{                    
  \inst{1} Faculty of Applied Physics and Mathematics, Gda\'nsk University of Technology, 80-233 Gda\'nsk, Poland\\
  \inst{2} Institute of Theoretical Physics and Astrophysics, University of Gda\'nsk, 80-952 Gda\'nsk, Poland \\
	\inst{3} National Quantum Information Centre in Gda\'nsk, 81-824 Sopot, Poland 
}

\abstract{
Classical objectivity as a property of quantum states---a view proposed to explain the observer-independent
character of our world  from quantum theory, is an important step in bridging the quantum-classical gap.
It was recently derived in terms of spectrum broadcast structures for small objects embedded in noisy photon-like environments. 
However, two fundamental problems have arisen: a description of objective motion and applicability to 
other types of environments. Here we derive an example of objective states of motion in quantum mechanics by showing
a formation of dynamical spectrum broadcast structures in the celebrated, realistic  model of decoherence---Quantum Brownian Motion. 
We do it for realistic, thermal environments and show their noise-robustness.
This opens a potentially new method of studying quantum-to-classical transition.}

\begin{document}

\maketitle

\section{Introduction}
Reconciliation of quantum theory with the classical world of everyday experience has been one of the central problems 
in our understanding of Nature \cite{Bohr,decoh}, touching such deep questions as is there any 'reality' out there \cite{Fine}. 
One of its aspects has been how to explain the objective character
of our world with fragile quantum systems, inevitably disturbed by measurements.
As quantum state is to date our most fundamental description of Nature, it is natural to look
for an explanation at this level. 
Indeed, recently specific quantum state structures---\emph{spectrum broadcast structures (SBS)}  \cite{object,sfera}, 
have been identified as responsible for the perceived objectivity, suggesting that the latter is, in fact, a property of
quantum states.
Building on the \emph{quantum Darwinism} idea \cite{ZurekNature,decoh}---a realistic form of decoherence theory
\cite{decoh} where the system of interest $S$ interacts with multiple environments $E_1,\dots,E_N$
and observers acquire information about $S$ through them,
it has been shown in \cite{object} (see also \cite{generic}) in a model- and dynamics-independent way
that the only, in a certain sense, states that encode objective states of the system are precisely the SBS: 
\ben\label{br2}
&&\varrho_{S:fE}=\sum_i p_i \ket{x_i}\bra{x_i}\otimes \varrho^{E_1}_i\otimes\cdots\otimes \varrho^{E_{fN}}_i,\\
&&\varrho^{E_k}_i\varrho^{E_k}_{i'\ne i}=0,
\een{equation}
where $fE$ is the observed portion of the environment $E$, $\{\ket{x_i}\}$ a pointer basis \cite{ZurekPRD}, $p_i$ pointer probabilities, and 
$\varrho^{E_1}_i,\dots,\varrho^{E_{fN}}_i$ some states of  $E_1,\dots,E_{fN}$
with orthogonal supports. As it is easy to see from (\ref{br2}), by properly measuring their portions of the environment (projecting on the supports
of $\varrho^{E_k}_i$), all the observers will obtain the same result $i$ without disturbing neither the system $S$ nor each other.
Since "seeing the same by many" without disturbance arguably defines a form of objectivity \cite{ZurekNature,object},
the states $\ket{x_i}$ become thus objective in this sense. 
Our approach is of course connected to the earlier studies  based on information redundancy \cite{ZurekNature}, but here 
we show it directly at the fundamental level of 
states, rather than using information-theoretical conditions, known so far only to be necessary \cite{object}.
A process of formation of a SBS \cite{sfera} is a weaker form \cite{my} of quantum state broadcasting \cite{broadcasting,CQ}.

A question now arises if such structures are indeed formed in realistic models of decoherence. 
Recently \cite{sfera}, their formation was
shown in the emblematic model of decoherence with scattering-type interactions: A small dielectric sphere illuminated by photons, but the resulting broadcast structure, and hence the objective states,
were static (described a fixed position) as the central system had no self-dynamics.
In this work we  study a fully dynamical model where both the system and the environment have own dynamics and
report a formation of objectively existing states of motion 
for the fundamental to all physics class of harmonic interactions.
In one of the universal models of decoherence---Quantum Brownian Motion (QBM) \cite{decoh,Ullersma,Petruccione}, which describes a central oscillator 
$S$ linearly coupled to a bath $E$ of oscillators, we show a formation, in the massive central system limit, of  novel dynamical spectrum broadcast structures (\ref{br2})
with time-evolving pointer states $\ket{x_i(t)}$. Due to developed correlations, information about this evolution is redundantly encoded in the environment
(in time-evolving, mutually orthogonal states $\varrho^{E_k}_i(t)$), even if the environment is noisy,
and in this sense it becomes objective \cite{ZurekNature,object}.
We model the noise as a thermal noise (with a ramification to arbitrary single-mode Gaussian noise) and numerically study
the effect as a function of the temperature, showing a certain noise-robustness. 
Surprisingly, in spite of being probably the most studied model of decoherence for decades \cite{decoh,Ullersma,Petruccione},
these state structures have not been noticed before (in the previous studies \cite{qbm_Zurek,Augusto} information-theoretical conditions were used, known so far to be only
necessary with their sufficiency being open \cite{object} and the environment was pure). Moreover, in contrast to the standard approaches 
\cite{decoh,Petruccione,qbm_Zurek}, we do not use the continuous approximation of the environment, keeping it discrete, thus deriving
objectivity in a more fundamental setup.

\section{The model}\label{model}
The central system $S$ is  a harmonic oscillator of a mass $M$ and a frequency $\Omega$, 
linearly coupled to the environment $E$---a bath of $N$ oscillators, each of a mass $m_k$ and a frequency $\omega_k$, $k=1,\dots, N$. 
The total Hamiltonian is \cite{decoh,Petruccione}:
\ben\label{H}
&&\hat H=\frac{\hat P^2}{2M}+\frac{M\Omega^2\hat X^2}{2}+\sum_{k=1}^N\left(\frac{\hat p_k^2}{2m_k}  
+\frac{m_k\omega_k^2\hat x_k^2}{2}\right)+\nonumber \\&&+\hat X\sum_{k=1}^N C_k\hat x_k,
\een
in the units $\hbar=1$; $\hat X, \hat P$ are the system's variables, $\hat x_k, \hat p_k$ describe the $k$-th environmental oscillator, 
and $C_k$ are the coupling constants.
The system's self-Hamiltonian we denote by $\hat H_S$, while the $k$-th environmental by $\hat H_k$. 
Our central interest is the information transfer from the system to the environment.
We will assume \cite{qbm_Zurek} that the central system is very massive, so it is effectively macroscopic, and will neglect 
all the back-reaction of the environment (non-dissipative regime). We note that this is exactly the opposite regime than the one 
used in the more familiar Born-Markov approximation and quantum master equation approaches to QBM \cite{decoh}.

Unlike in the usual approaches \cite{qbm_Zurek,Petruccione}, we also do not pass here to the continuous limit and to 
a continuous spectral density function, working all the time with the discrete environment. 
To make the decoherence possible, we assume a random distribution of $\omega_k$'s (cf. \cite{Zurek_spins}).
This choice is  some form of a spectral density, but we keep it discrete. 
Furthermore, we will work in the off-resonant regime:
\begin{equation} \label{offres}
\omega_k\ll\Omega\ \text{ or }\ \omega_k\gg\Omega\quad \text{for all} \ k,
\end{equation}
so that, as will become clear later, a single environmental oscillator alone will not decohere the central system \cite{qbm_Zurek,Augusto,Petruccione}.
Albeit possible, that would be a somewhat trivial situation as we are interested here in a regime where a single environment 
carries a vanishingly small amount of information about the system
\cite{comm}. We will thus study collective effects and following \cite{object,sfera}, we will group
the environments into macro-fractions---fragments scaling with the total number of oscillators $N$, and study their
information content. 

\section{The dynamics}\label{dynamics}
Although the exact solution of the model is possible as the Hamiltonian (\ref{H}) is quadratic \cite{Ullersma},
for the purpose of this study we will use the  approximate method of Refs. \cite{qbm_Zurek,Augusto} taking advantage
of the assumed high mass of the central system
(a type of a non-adiabatic Born-Oppenheimer approximation with classical trajectories; see e.g. \cite{NBO}).
In this approximation, the system $S$ evolves according to its self-Hamiltonian $\hat H_S$, with this evolution further approximated using
classical trajectories $X(t;X_0)$, while the environment is driven along each of this trajectory. The resulting state is:  
\begin{equation}\label{final}
\ket{\Psi_{S:E}}=\int dX_0 \phi_0(X_0) e^{-i\hat H_St}|X_0\rangle\otimes \hat U_{E}(X(t;X_0))\ket{\psi_0},
\end{equation} 
$\hat U_{E}(X(t;X_0))$ is the evolution generated by $\hat H_{E}(X)\equiv\sum_k(\hat H_k+C_kX\hat x_k)$ for the trajectory $X(t;X_0)$ and $\ket{\phi_0}$,$\ket{\psi_0}$
are initial states of $S$ and $E$ respectively. Formally, (\ref{final}) is obtained by a controlled-unitary evolution \cite{sfera}:
\begin{equation}\label{USE}
\hat U_{S:E}(t)=\int dX_0 e^{-i\hat H_St}\ket{X_0}\bra{X_0}\otimes \hat U_{E}(X(t;X_0)),
\end{equation}
acting on the initial state $\ket{\phi_0}\ket{\psi_0}$. Since $\hat H_S$ is quadratic, the trajectory approximation is actually
exact  (the semi-classical propagator is exact). For simplicity,  we will limit ourselves to
trajectories obtained when the system is initially in the squeezed vacuum state (cf. \cite{qbm_Zurek,Augusto}): 
$\ket{\phi_0}=\hat S(r)\ket{0}$, where $\hat S(r)\equiv e^{r(\hat a ^2-\hat a^{\dagger 2})/2}$.
Especially interesting is a highly momentum squeezed state due to its large coherences in the position.
We may than assume that the initial velocity of each trajectory is zero so that
$X(t;X_0)=X_0\cos(\Omega t)$. The analysis of the high initial position squeezing, for which $X_0=0$ and $X(t;X_0)=X_0\sin(\Omega t)$, 
will be analogous.  We solve for $\hat U_{E}(X(t;X_0))$ using
$\hat U_{E}(X(t;X_0))=\lim_{n\to\infty} \big(\prod_{r=1}^n \exp[-i\hat H_{E}(t_r)\Delta t]\big)$, $\Delta t\equiv t/n$, $t_r\equiv r\Delta t$ and obtain:
\begin{eqnarray}
&&\hat U_{E}(X(t;X_0))=\bigotimes_{k=1}^N \hat U_k(X_0;t),\label{Ue}\\
&&\hat U_k(X_0;t)\equiv e^{i\zeta_k(t)X_0^2}e^{-i\hat H_k t}\hat D\left(\alpha_k(t)X_0\right),\label{UI}
\end{eqnarray}
so that (\ref{USE}) has the following form:
\begin{eqnarray}\label{USE2}
\hat U_{S:E}(t)=&&e^{-i\hat H_St}\otimes e^{-i\sum_k\hat H_k t}\times\\
&&\times\int dX_0 \ket{X_0}\bra{X_0}\otimes e^{i\zeta_k(t)X_0^2}\hat D\left(\alpha_k(t)X_0\right).\nonumber
\end{eqnarray}
Here $\hat D(\alpha)\equiv e^{\alpha \hat a^\dagger-\alpha^*\hat a}$ is the displacement operator \cite{Perelomov}, $\hat a^\dagger,\hat a$ are the
creation and annihilation operators, 
$\zeta_k(t)$ is a dynamical phase (as we will show irrelevant for our calculations), and: 
\begin{eqnarray}
\alpha_k(t)\equiv -\frac{C_k}{2\sqrt{2m_k\omega_k}}\left[\frac{e^{i(\omega_k+\Omega)t}-1}{\omega_k+\Omega}+
\frac{e^{i(\omega_k-\Omega)t}-1}{\omega_k-\Omega}\right]\label{ak}
\end{eqnarray}
for the momentum squeezing and: 
\begin{eqnarray}
\alpha_k(t)\equiv -\frac{C_k}{2i\sqrt{2m_k\omega_k}}\left[\frac{e^{i(\omega_k+\Omega)t}-1}{\omega_k+\Omega}-
\frac{e^{i(\omega_k-\Omega)t}-1}{\omega_k-\Omega}\right].\label{ak_p}
\end{eqnarray}
for the position squeezing.

\section{Dynamical Spectrum Broadcast Structure}

The formation of SBS (\ref{br2}) is equivalent to: 
(i) decoherence and (ii) perfect disntinguishability of post-interaction environmental states \cite{sfera}. 
We study the evolved $S:E$ state under the approximations described in the previous Section and after
tracing over a fraction $(1-f)E$, $f\in(0,1)$, of the environment 
that passes unobserved and is necessary for the decoherence: 
$\varrho_{S:fE}(t)\equiv tr_{(1-f)E}\varrho_{S:E}(t)$,
$\varrho_{S:E}(t)\equiv \hat U_{S:E}(t)(\ket{\phi_0}\bra{\phi_0}\otimes\bigotimes_k\varrho_{0k})\hat U_{S:E}(t)^\dagger$.
We assume the environment to be initially in a thermal state so that all $\varrho_{0k}$'s are thermal states with the same temperature $T$ 
(later we will generalize to arbitrary single-mode Gaussian states).
Although (\ref{USE}) is formally written 
with a continuous distribution of $X_0$, it in fact stands for a limit of  finite divisions $\{\Delta_i\}$ of the real line $\mathbb R$, with 
$\ket{X_0} \bra{X_0}$ approximated by orthogonal projectors $\hat \Pi_{\Delta}$ on the intervals $\Delta$ (see e.g. \cite{GalindoPascual}). 
From (\ref{USE}-\ref{UI}) we obtain:
\ben
&&\nonumber \varrho_{S:fE}(t)=\sum_\Delta e^{-i\hat H_{S}t}\hat \Pi_\Delta \ket{\phi_0}\bra{\phi_0} \hat \Pi_\Delta e^{i\hat H_{S}t}\bigotimes_{k=1}^{fN} \varrho_k(X_\Delta;t)\\
&& +\sum_{\Delta\ne\Delta'} \Gamma_{X_\Delta,X_{\Delta'}}(t) e^{-i\hat H_St}\hat \Pi_\Delta \ket{\phi_0}\bra{\phi_0} \hat \Pi_{\Delta'}e^{i\hat H_St} \label{mama} \\ 
&&\otimes\bigotimes_{k=1}^{fN} \hat U_k(X_\Delta;t) \varrho_{0k}\hat U_k(X_{\Delta'};t)^\dagger,\nonumber
\een
where $fN$ denotes the number of observed oscillators, $X_\Delta$ is some position within $\Delta$, and:
\begin{eqnarray}
\label{rI}
&&\varrho_k(X;t)\equiv \hat U_k(X;t) \varrho_{0k}\hat U_k(X;t)^\dagger,\\
&&\Gamma_{X,X'}(t)\equiv \prod_{k\in (1-f)E}tr\left[\hat U_k(X;t) \varrho_{0k}\hat U_k(X';t)^\dagger\right],\label{G}
\end{eqnarray}
the latter being the decoherence factor due to the traced fraction $(1-f)E$ of the environment (for compactness
we denote the system's initial position by $X$ rather than $X_0$). It governs vanishing of the off-diagonal part in (\ref{mama}) in the trace-norm
\cite{sfera}. A closed formula for $|\Gamma_{X,X'}(t)|$ for general initial states $\varrho_{0k}$ is possible, using the fact \cite{Prep} that one can always write 
$\varrho_{0k}=(1/\pi)\int d^2\alpha P_k(\alpha) \ket\alpha\bra\alpha$, where $\ket \alpha$ are the usual coherent states \cite{Perelomov}
and $P_k(\alpha)$ is in general a distributional Glauber-Sudarshan $P$-representation: 
\begin{eqnarray}
&&\left| \Gamma_{X,X'}(t)\right|=\prod_{k\in (1-f)E}e^{-\frac{|\alpha_k(t)|^2}{2}(X-X')^2}\times\nonumber\\
&&\left|\int \frac{dqdp}{\pi}P_k(q,p)e^{2i(X-X')\left[q\text{Im}\alpha_k(t)-p\text{Re}\alpha_k(t)\right]}\right|.
\end{eqnarray}
(phases $\zeta_k(t)$, cf. (\ref{UI}), cancel due to the modulus). Here:
\begin{eqnarray}
|\alpha_k(t)|^2&=&\frac{C_k^2\omega_k}{2m_k(\omega_k^2-\Omega^2)^2}
\bigg[\left(\cos\omega_kt-\cos\Omega t\right)^2\nonumber\\
&&+\left(\sin\omega_kt-\frac{\Omega}{\omega_k}\sin\Omega t\right)^2\bigg]\label{aT}
\end{eqnarray}
for an initial momentum squeezed state of $S$  (cf. (\ref{ak})).
For thermal states at temperature $T$, $P_k(q,p)=(1/\bar n_k)e^{-(q^2+p^2)/\bar n_k}$, $\bar n_k=1/(e^{\beta\omega_k}-1)$, $\beta\equiv1/k_BT$ and 
the corresponding 
decoherence factor is given by 
\cite{Petruccione}:
\ben\label{GT}
&&\left| \Gamma_{X,X'}(t)\right|=\nonumber \\ &&\prod_{k\in(1-f)E}\exp\left[-\frac{(X-X')^2}{2}|\alpha_k(t)|^2\text{cth}\left(\frac{\beta\omega_k}{2}\right)\right],
\een
where $\text{cth}(\cdot)$ is the hyperbolic cotangent. 
From (\ref{aT}) it is clear that bands near the resonant mode $\omega_k\approx \Omega$
would be enough to effectively decohere the system \cite{qbm_Zurek,Augusto}. But here we want to study the opposite, more subtle, regime
where a single mode has a very small influence on the system's coherence. This motivates the condition (\ref{offres}).
Due to discrete and random $\omega_k$'s,  $\left| \Gamma_{X,X'}(t)\right|$ 
is in our study an almost periodic function of time \cite{ap}. We analyze it later.

Next, we turn to the diagonal part in (\ref{mama}), reverting to the continuum limit. We group the observed 
environment $fE$ into $\mathcal M$ macro-fractions of an equal size of $fN/\mathcal M$ oscillators each \cite{object,sfera}
and show that there is a regime, where the 
states of each macro-fraction (cf. (\ref{rI}))
$\varrho_{mac}(X;t)\equiv\bigotimes_{k\in mac}\varrho_k(X;t)$
become perfectly distniguishable for different $X$ ($k\in mac$ means $k$ running through the oscillators in a given macro-fraction $mac$).
We use the generalized overlap \cite{Fuchs}: 
\be\label{B}
B(\varrho_1,\varrho_2)\equiv tr\sqrt{\sqrt{\varrho_1}\varrho_2\sqrt{\varrho_1}}
\ee
as the most convenient measure of distinguishability (cf. (\ref{br2})): 
$\varrho_1$ and $\varrho_2$ are perfectly distinguishable, $\varrho_1\varrho_2=0$, if and only if $B(\varrho_1,\varrho_2)=0$.  
A calculation for thermal $\varrho_{0k}$'s gives (see Appendix \ref{genoverlp}):
\be\label{BT}
B^{mac}_{X,X'}(t)=\prod_{k\in mac}\exp\left[-\frac{(X-X')^2}{2}|\alpha_k(t)|^2\text{th}\left(\frac{\beta\omega_k}{2}\right)\right],
\ee
where $B^{mac}_{X,X'}(t)\equiv B[\varrho_{mac}(X;t),\varrho_{mac}(X';t)]$ measures the distinguishability of the system's initial positions $X$, $X'$ as
recorded into macro-fractions.
Note, however, that  the states $\varrho_{mac}(X;t)$ depend not only on $X$, but on the whole classical motion through (\ref{USE}).
From (\ref{GT},\ref{BT}) $\lim_{T\to\infty}|\Gamma_{X,X'}(t)|=0$, i.e. hot environments decohere
the central system better, but as $\lim_{T\to\infty}B^{mac}_{X,X'}(t)=1$ 
they are unable to discriminate its positions, irrespectively of the observed macro-fraction 
size---hot environments are too noisy (the initial states $\varrho_{0k}$ are too close to the maximally mixed state)
to store any information  (cf. (\ref{rI})). Note that the factor $\text{th}(\beta\omega_k/2)$, appearing in both (\ref{GT},\ref{BT}), is nothing else but the purity $tr(\varrho_{0k}^2)$.

\begin{figure}[t]
        \centering
        \subfigure{
                \includegraphics[width=0.22\textwidth]{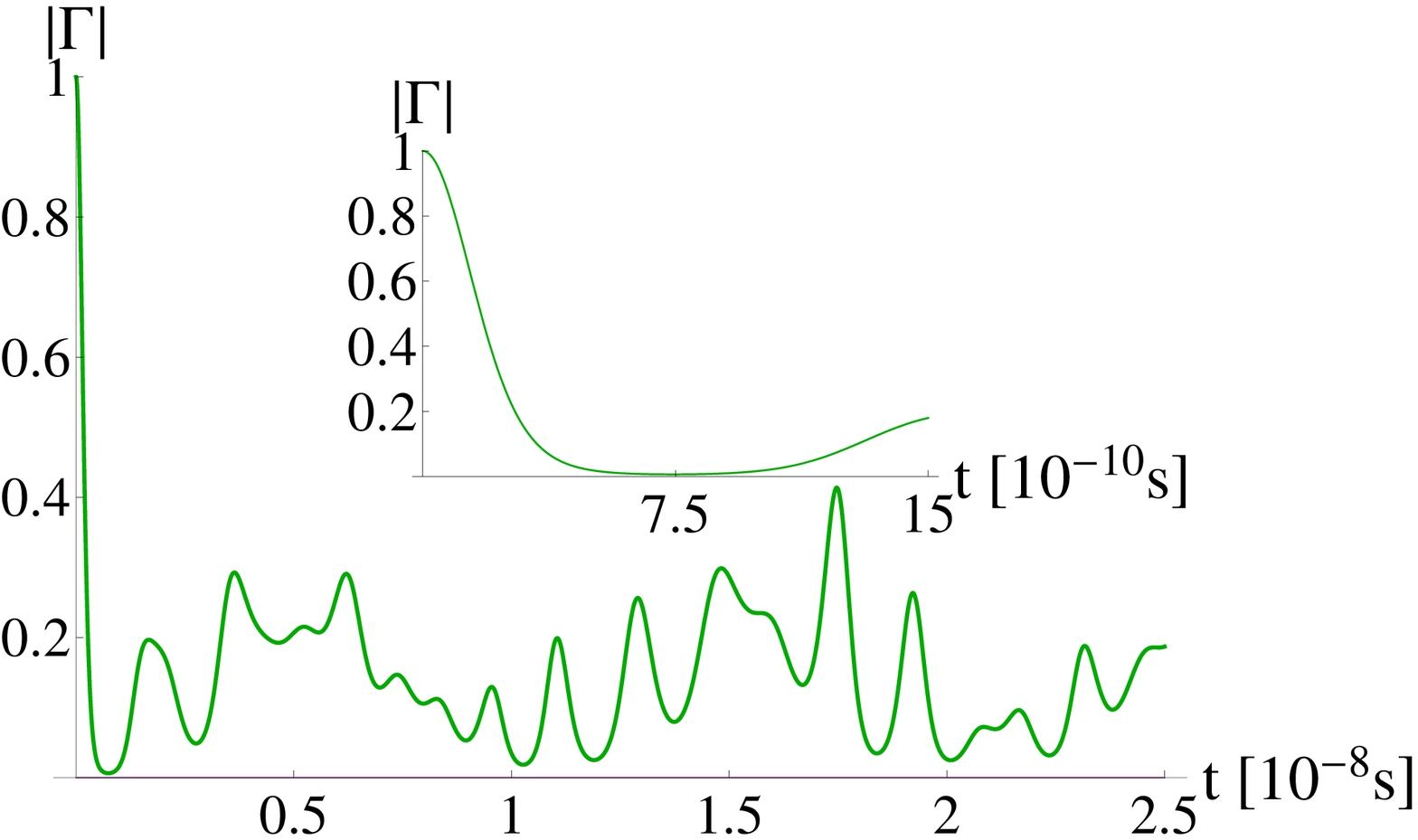}
} 
       \subfigure{
                \includegraphics[width=0.22\textwidth]{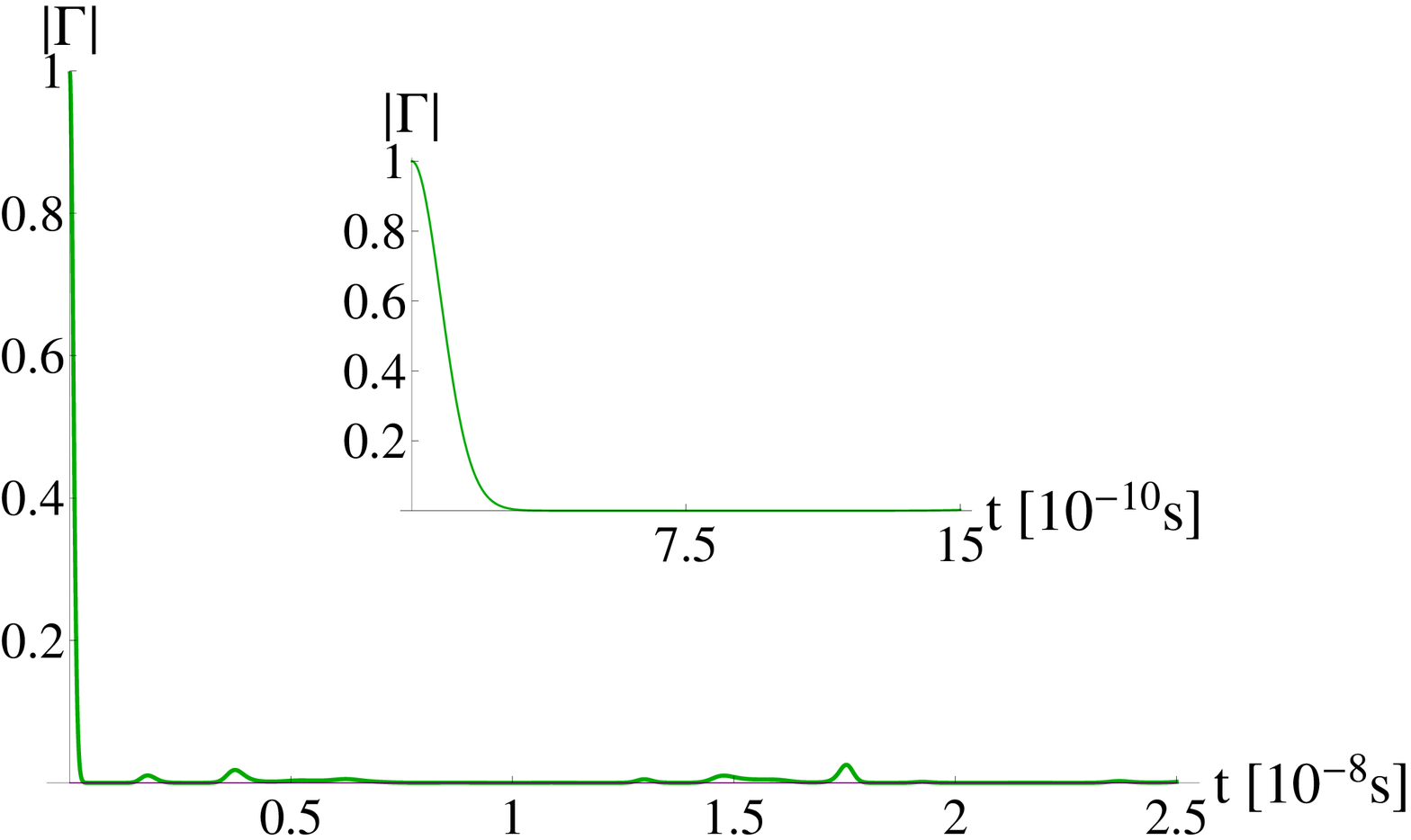}
}
	 \subfigure{
                \includegraphics[width=0.22\textwidth]{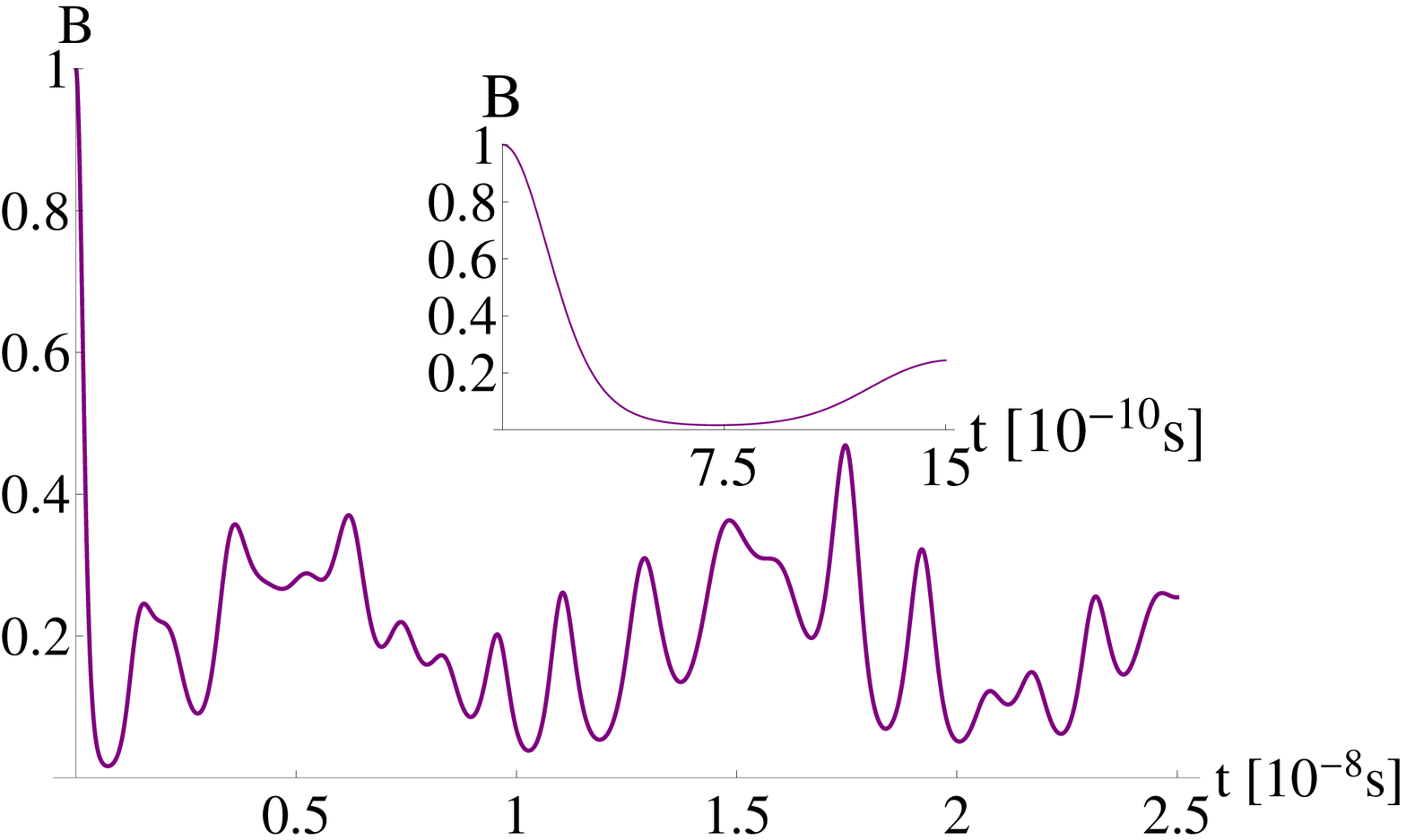}
}       
        \subfigure{
                \includegraphics[width=0.22\textwidth]{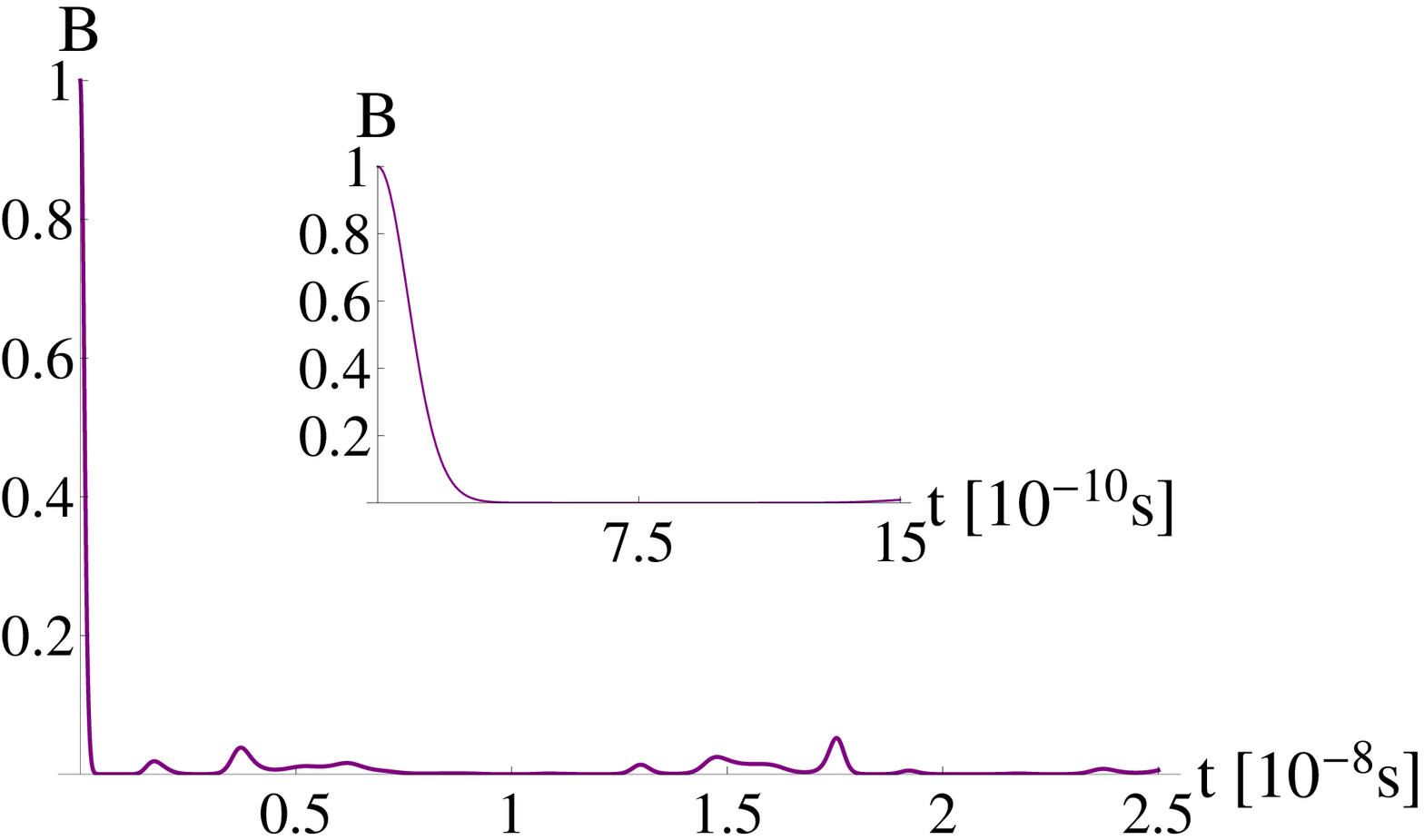}
}
\caption{(Color online) Time dependencies of $|\Gamma_{X,X'}(t)|$ (a),(b) 
and $B^{mac}_{X,X'}(t)$ (c),(d) for the system initially in a momentum squeezed state
for macro-fraction sizes $N=10$ (a),(c) and $N=30$ (b),(d) and $T=10^{-2}$K.
The inserts show  short-time behavior.}
\label{time}
\end{figure}

\section{Numerical analysis}
We first analyze the case when the system $S$ initially in a momentum squeezed state.
Both $|\Gamma_{X,X'}(t)|$ and $B^{mac}_{X,X'}(t)$ depend on the same almost periodic function
of time (\ref{aT}), too complicated for an immediate analytical study. 
In this work we analyze it numerically. 
We set: $M=10^{-5}$kg, $\Omega=3 \times 10^{8}$s$^{-1}$,
$\omega_k$'s independently, identically and uniformly distributed in the interval $3\dots 6 \times 10^{9}$s$^{-1}$ to satisfy (\ref{offres}), and $|X-X'|=10^{-9}$m. 
We assume that $C_k$ depend only on the masses: 
$C_k\equiv 2\sqrt{(M m_k \tilde\gamma_0 )/\pi}$, and $\tilde\gamma_0 =0.33\times 10^{18}$ s$^{-4}$ is a constant.
We assume a symmetric situation: The size of the traced macro-fraction $(1-f)E$ in (\ref{GT}) is the same as the size of 
the observed one $mac$ in (\ref{BT}).
Intuitively, for large enough macro-fractions for a given $T$, $|\Gamma_{X,X'}(t)|$ and $B^{mac}_{X,X'}(t)$ should decay rapidly and have small typical fluctuations due to the large 
amount of random phases in (\ref{aT}), indicating decoherence and perfect distingushability. 
This is confirmed in Fig.~\ref{time}. From Figs.~\ref{time}b,d we see that for 30 oscillators both functions decay rapidly, while
for 10 oscillators they do not---the macro-fraction is too small for the given $T$. 

\begin{figure}[t]
\centering

 \subfigure{\includegraphics[scale=0.25]{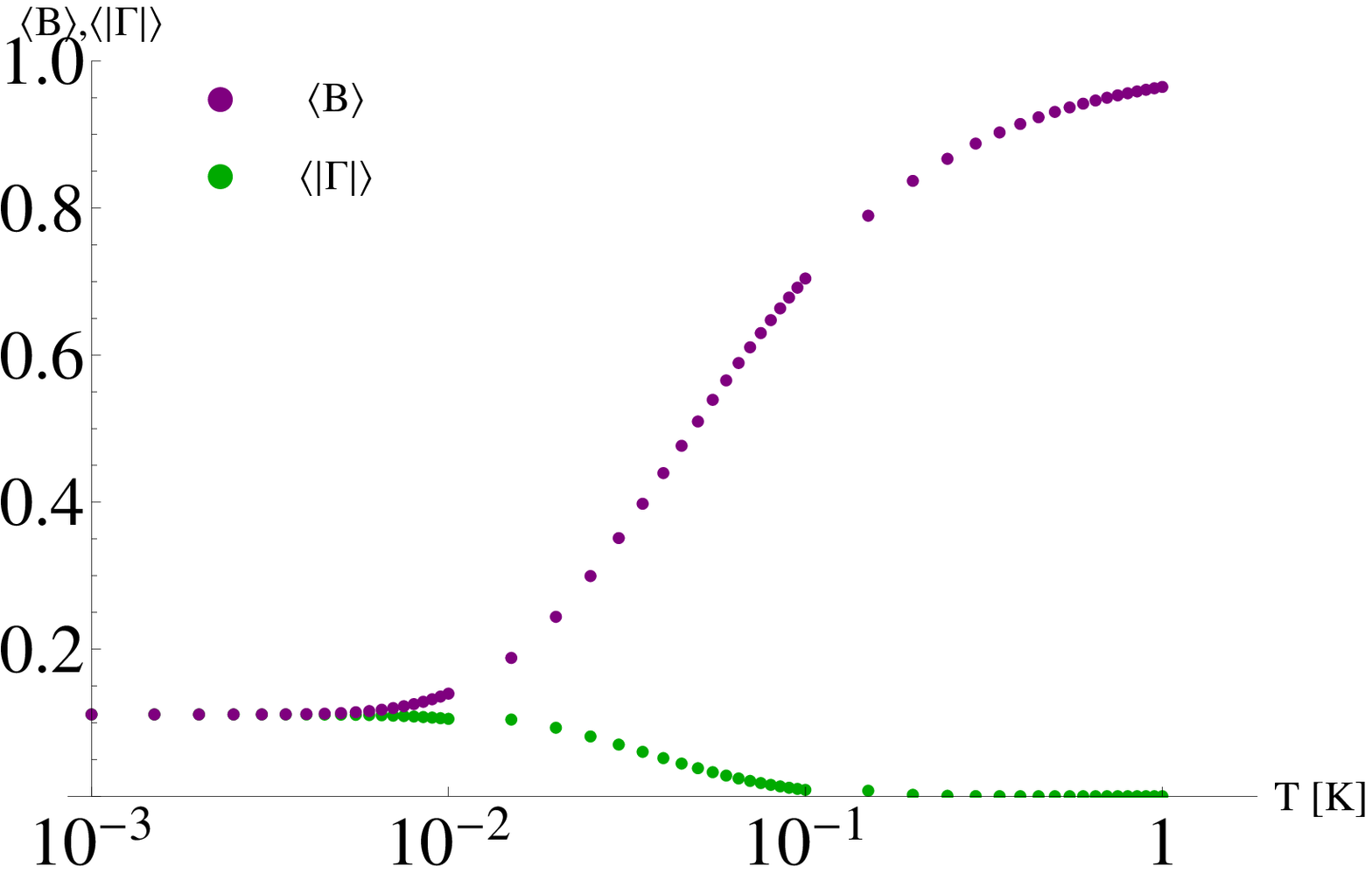}
}
 \subfigure{\includegraphics[scale=0.25]{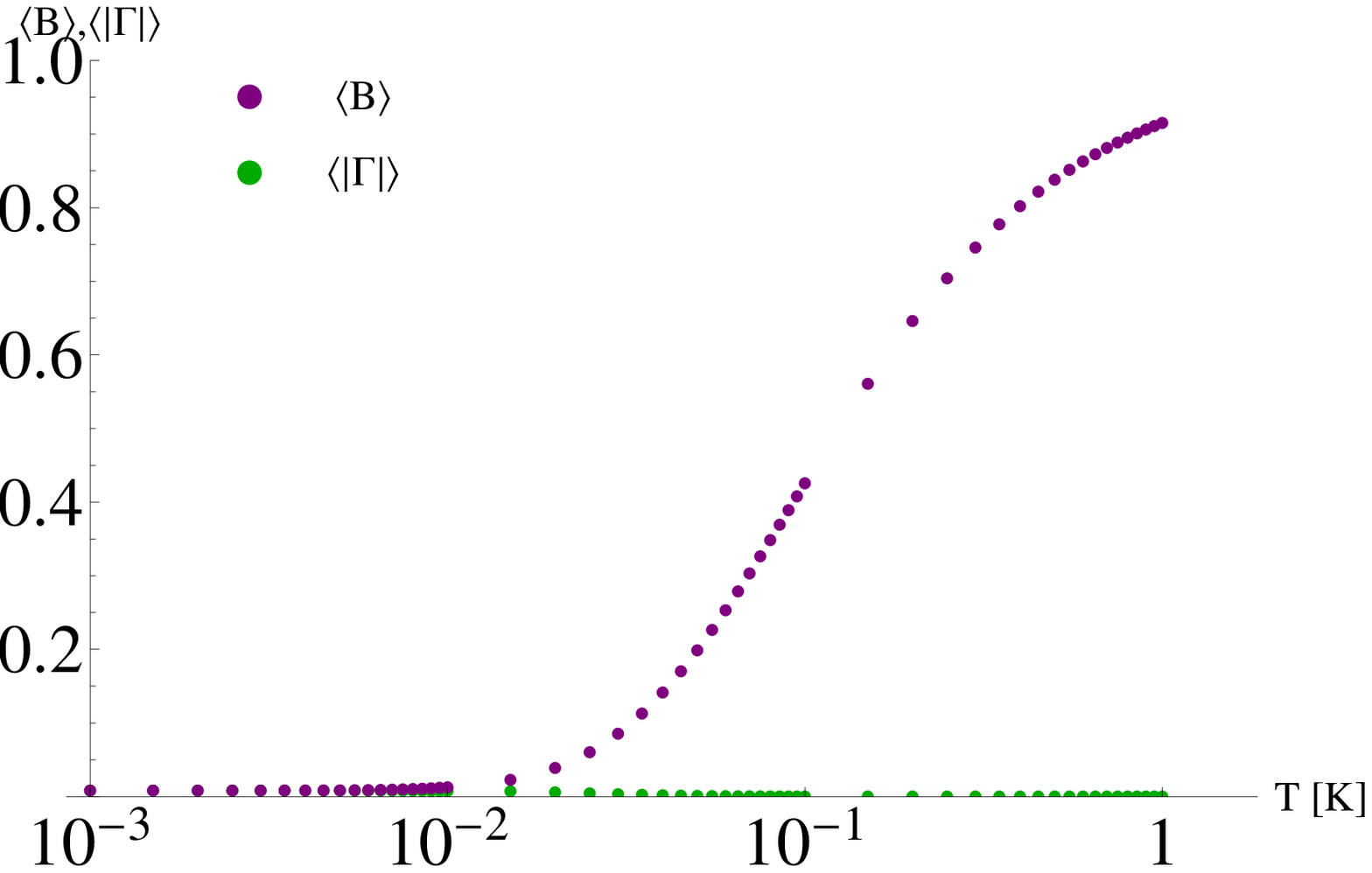}
}
\caption{(Color online) Time-averaged  $|\Gamma_{X,X'}|$ (lower traces) and 
$B^{mac}_{X,X'}$ (upper traces) for the system initially in a momentum squeezed state
as functions of the temperature (on the logarithmic scale) for macro-fraction sizes  $N=10$ (a) and $N=30$ (b).
Plot (b) shows formation of the broadcast state for $T<10^{-2}K$.}\label{av}
\end{figure}

We further analyze, Fig.~\ref{av}, the time averages 
$\left\langle |\Gamma_{X,X'}| \right\rangle = (1/\tau) \int^{\tau}_{0} dt |\Gamma_{X,X'}(t)|$, 
$\left\langle B^{mac}_{X,X'}\right\rangle = (1/\tau) \int^{\tau}_{0} dt B^{mac}_{X,X'}(t)$ as functions of the temperature $T$
with $\tau$ taken large ($\sim 1s$): Since both functions are non-negative, vanishing of their time averages is a good indicator of the functions 
having small typical fluctuations above zero. From Fig.~\ref{av}a one sees that, in the chosen parameter range, 
there is no formation of the broadcast state for a macro-fraction of $10$ oscillators: While 
$\left\langle |\Gamma_{X,X'}| \right\rangle\approx 0$  
(the lower trace) 
for $T\approx 10^{-1}K$, $\left\langle B^{mac}_{X,X'}\right\rangle\approx 0.6$ (the upper trace).
The state decoheres, but at too high a temperature 
to store a perfect record of the system's position.
From (\ref{mama}), the post-interaction partial state is then of a, so called, Classical-Quantum (CQ) type \cite{CQ}.
However, increasing the size to $30$ oscillators both traces become practically zero up to $T\approx 10^{-2} K$, 
as one sees from Fig.~\ref{av}b (cf. Fig.~\ref{time}b,d). 
This serves as a numerical evidence of a formation of the spectrum broadcast structure (\ref{br2}), and hence objectivisation \cite{object},
in the Quantum Brownian Motion model with a massive central system, initiated in a highly momentum-squeezed state, i.e. 
possessing large coherences in the position. This is our main result.

\begin{figure}[t]
        \centering
        \subfigure{
               \includegraphics[width=0.19\textwidth]{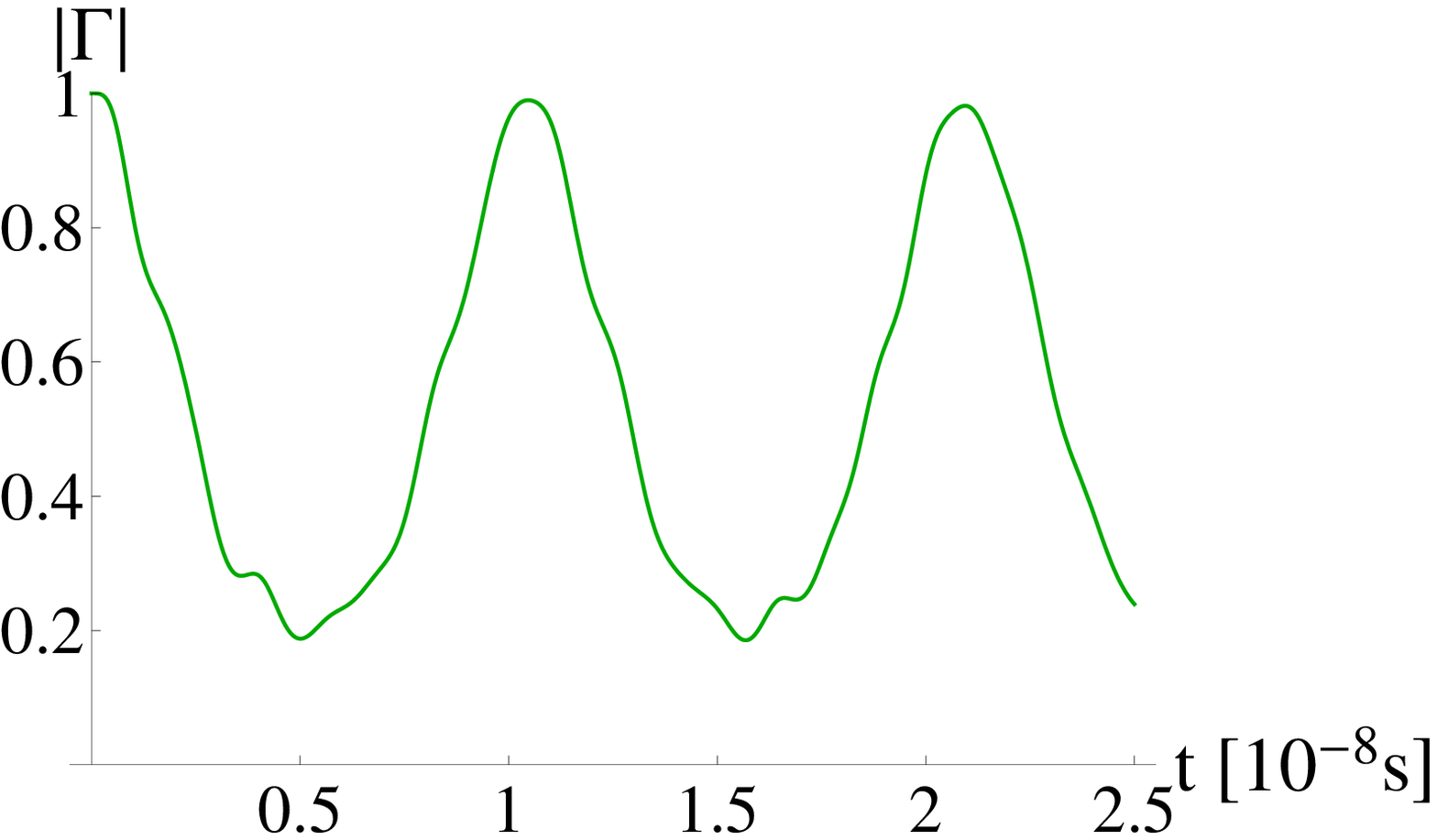}
} 
        \subfigure{
                \includegraphics[width=0.19\textwidth]{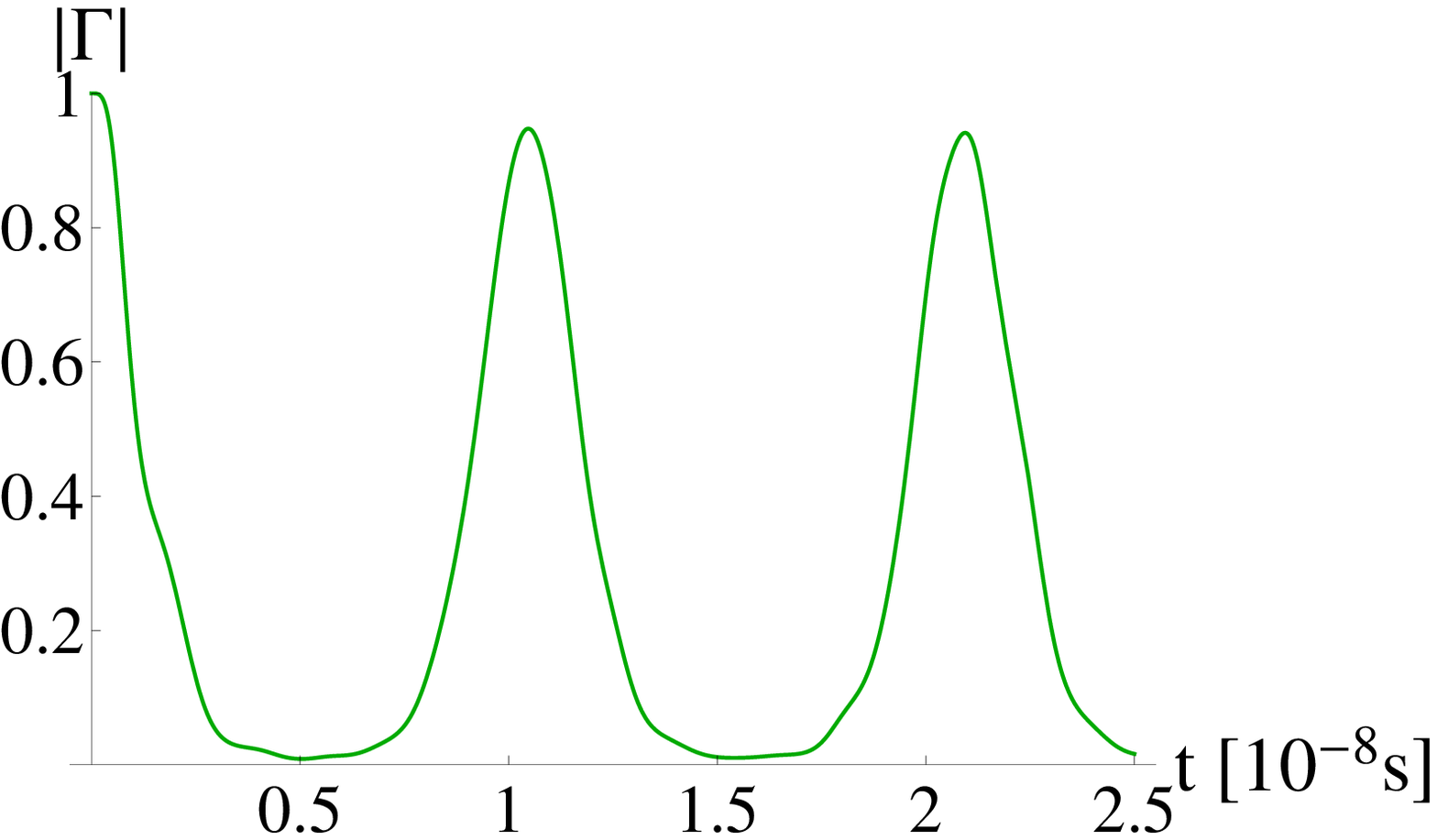}
}
	 \subfigure{
                \includegraphics[width=0.19\textwidth]{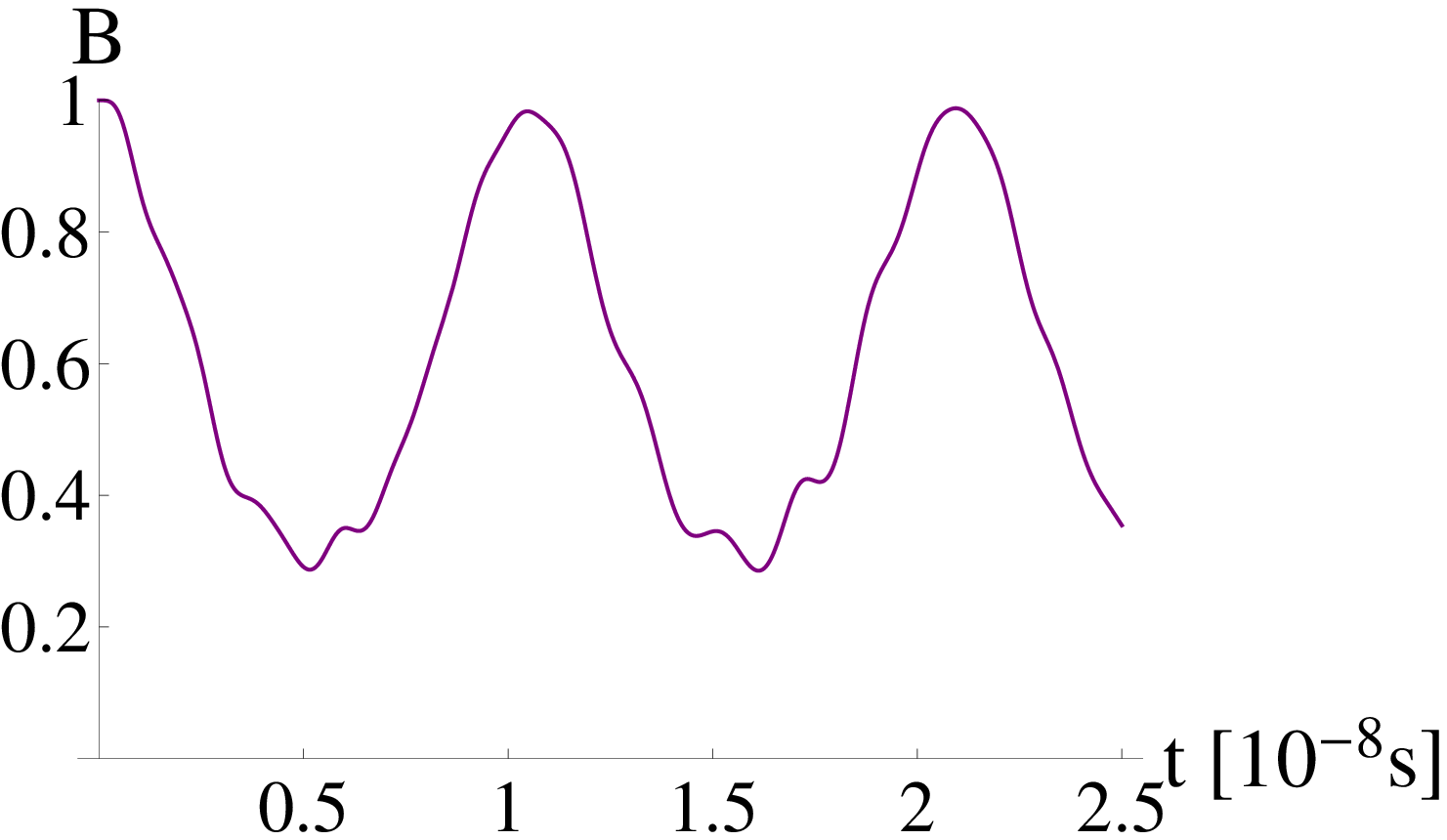}
}       
        \subfigure{
                \includegraphics[width=0.19\textwidth]{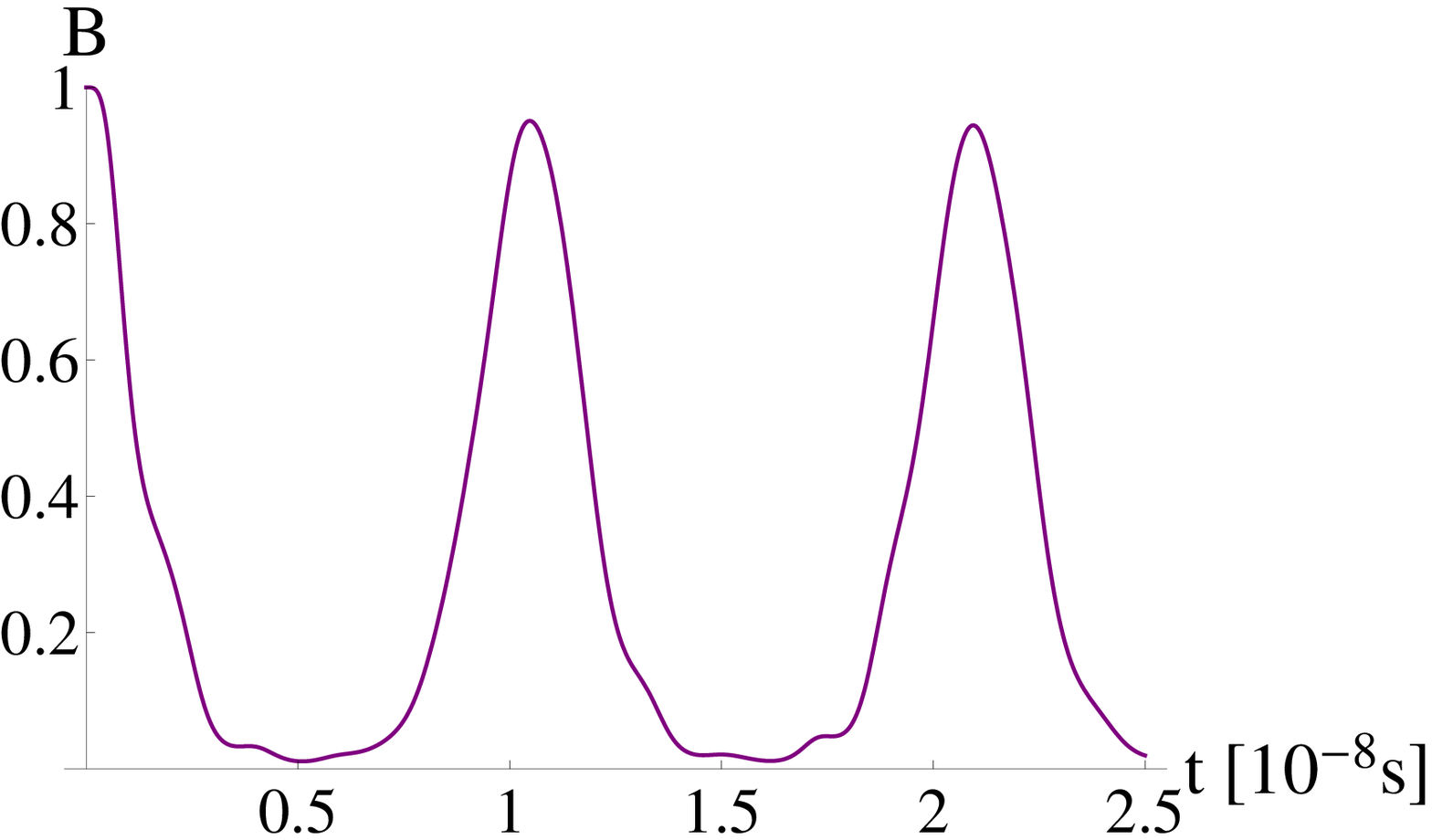}
}
\caption{(Color online) Time dependencies of $|\Gamma_{X,X'}(t)|$ (a),(b) 
and $B^{mac}_{X,X'}(t)$ (c),(d) for the system initially in a position squeezed state
for  macro-fraction sizes $N=10$ (a),(c) and $N=30$ (b),(d) and in the same temperature $T=10^{-2}$K as in Fig.~\ref{time}.
At this temperature the functions differ little.
}
\label{time_p}
\end{figure}

The situation with initial position squeezing, for which the trajectories are given by $X(t;X_0)=X_0\sin(\Omega t)$
is quite different. Under exactly the same conditions as above there is no decoherenece neither orthogonalization 
for macrofractions of both 10 and 30 oscillators as Fig.~\ref{time_p} shows. Actually the plots suggest that both functions
are periodic in time (even increasing th macrofraction size to 100), so there is a periodic revival of coherence. 
This in general agrees with the findings of \cite{Augusto}.

\section{Dynamical Objectivity}Let us assume that a SBS is formed, i.e. both $|\Gamma_{X,X'}(t)|$ and $B^{mac}_{X,X'}(t)$
approach zero. Then from (\ref{mama}) (taking the usual continuum limit of the sum):
\begin{eqnarray}
&&\varrho_{S:fE}(t)=\int dX_0 \left|\langle X_0|\phi_0\rangle\right|^2 \times\label{DB}\\
&& \times\ket{X(t)}\bra{X(t)} 
\otimes\varrho_{mac_1}(X_0;t)\otimes \cdots\otimes \varrho_{mac_{\mathcal M}}(X_0;t),\nonumber
\end{eqnarray}
where $\ket{X(t)}\equiv e^{-i\hat H_St}\ket {X_0}$, we have grouped $fE$ into $\mathcal M$ macro-fractions
and $\varrho_{mac_i}(X_0;t)$ have orthogonal supports (for large enough $t$; cf. e.g.  Fig.~\ref{time}d).
What appears in (\ref{DB}) is a novel, compared to the previous studies \cite{object,sfera}, 
dynamical spectrum broadcast structure (dSBS).
Because the system now has its own dynamics, the pointers $\ket{X(t)}$ are now  states of motion---they 
evolve on a time-scale $t_S\sim 2\pi/\Omega$, rather than being static as in \cite{sfera},
and a time-dependent SBS is formed with a reference to these evolving pointers.
For the example studied in the previous Section, the respective time-scales are
$t_S\sim 2\times10^{-8}s$ and from Fig.~\ref{time}b,d $t_{SBS}\sim 2\times 10^{-10}s$ 
so that the SBS is formed two orders of magnitude faster than the intrinsic system evolution.
Thanks to it, all the observers will measure the same initial position (= the oscillation amplitude) $X_0$, 
leaving the (by now decohered) system undisturbed in its state of motion.
But the traces of this motion  are present
in the environment not only through $X_0$---each state $\varrho_{mac}(X_0;t)$ 
depends on the whole trajectory $X(t;X_0)$ (cf. (\ref{final})).

The intuitive picture is that while the system rotates on its intrinsic timescale, the environment
follows this movement and past the transient period a spectrum broadcast structure is being continuously formed, leading 
to a perception of objective position at each moment of time.
Of course due to the neglected back reaction on the system, the structure (\ref{DB}) 
is only a first approximation to this situation, as e.g. there is no dynamical production of coherences in the system's position.
The next logical step would be to include the back reaction. 

\section{General Gaussian Initial States}
We recall \cite{gaus} that an arbitrary single-mode Gaussian state can be parametrized as follows:
$
\varrho=e^{i\psi \hat a^\dagger \hat a}\hat D(\gamma)\hat S(\xi)\varrho_T\hat S(\xi)^\dagger \hat D(\gamma)^\dagger e^{-i\psi \hat a^\dagger \hat a},
$
where $\hat S(\xi)\equiv e^{(\xi^*\hat a ^2-\xi \hat a^{\dagger 2})/2}$, $\xi\equiv re^{i\theta}$, and
$\varrho_T$ is some thermal state. Parametrizing each $\varrho_{0k}$ as above 
leads to the same expressions (\ref{GT},\ref{BT}) but with $\alpha_k(t)$ (cf. (\ref{ak})) substituted by:
$
\tilde\alpha_k(t)\equiv \text{ch} r\left[e^{-i\psi} \alpha_k(t) - e^{i(\psi+\theta)} \alpha_k(t)^*\text{th} r\right].
$
Introduction of a squeezing  increases the temperature range 
where a dynamical  SBS can be formed via increasing the informational capacity of the environment,
e.g. for $r=5$, the temperature range is increased up to $T=1K$ \cite{Photonics}.

\section{Concluding remarks}
Our findings generally agree with that of \cite{qbm_Zurek,Augusto} in that there is a parameter range in QBM such that objectivity appears, 
but it has been obtained with a deeper analysis directly on quantum states, uncovering previously unnoticed dynamical 
spectrum broadcast structures. Our method, although
developed here in a specific model, is in fact much more universal and can be generalized to test other decoherence models
for a presence of dynamical forms of objectivity: One checks if states of the type (\ref{DB}) are formed during the evolution.
One immediate generalization is to allow for other trapping potentials than harmonic and other couplings than linear (see e.g. \cite{Pietro}). 
Another, is to study finite-dimensional systems, e.g. spins \cite{spins},
but a far more challenging generalization would be an application to quantum fields,
leading to objective dynamical classical fields.
Finally, a possible connection between Markovianity/non-Markovianity of the evolution and a formation of broadcast structures can also be studied 
\cite{Galve}.
\section{The generalized overlap $B^{mac}_{X,X'}(t)$ for thermal environment states}\label{genoverlp}

We calculate 
\be\label{Bm}
B^{mac}_{X,X'}(t)\equiv B[\varrho_{mac}(X;t),\varrho_{mac}(X';t)],
\ee
for 
$\varrho_{mac}(X;t)\equiv\bigotimes_{k\in mac} \hat U_k(X;t)\varrho_{0k}\hat U_k(X;t)^\dagger$ and $\varrho_{0k}$  thermal.
The above distinguishability  measure  \cite{Fuchs} scales with the
tensor product $B\big(\bigotimes_k\varrho_k,\bigotimes_k\varrho'_k\big)=\prod_k B(\varrho_k,\varrho'_k)$,
so that it is enough to calculate it for a single environment. Dropping the explicit dependence
on  $k$ and denoting a single-system overlap by  $B^{mic}_{X,X'}(t)$ we obtain:
$B^{mic}_{X,X'}(t)= tr\sqrt{\sqrt{\varrho_0}\hat U(X';t)^\dagger\hat U(X;t)\varrho_0\hat U(X;t)^\dagger\hat U(X';t)\sqrt{\varrho_0}}$,
where we have pulled the extreme left and right unitaries out of the both square roots and used the cyclic property of the trace to cancel them out.
Thus, modulo phase factors:
$\hat U(X';t)^\dagger\hat U(X;t) \simeq\hat D\left(\alpha(t)(X-X')\right)\equiv\hat D(\eta_t)$,

Next, assuming all the $\varrho_{0k}$ are thermal with the same temperature, 
we use the $P$-representation for the middle $\varrho_0$ under the square root in $B^{mic}_{X,X'}(t)$:
$\varrho_0=\int d^2\gamma /(\pi\bar n)\exp\left(-|\gamma|^2/\bar n\right)\ket\gamma\bra\gamma$, 
where $\bar n=1/(e^{\beta\omega}-1)$, $\beta\equiv1/k_BT$. 
Denoting the Hermitian operator under the square root in $B^{mic}_{X,X'}(t)$ by $\hat A_t$, we obtain:
$\hat A_t 
=\int d^2 \gamma/ (\pi\bar n)e^{-|\gamma|^2/\bar n}\sqrt{\varrho_0}\ket{\gamma+\eta_t}\bra{\gamma+\eta_t}\sqrt{\varrho_0}$.
To perform the square roots above we use the Fock representation:
$\varrho_0=\sum_n\left( \bar{n}^n/(\bar n +1)^{n+1} \right)\ket n\bra n$, so that:
\begin{eqnarray}
\hat A_t&=&\int\frac{d^2\gamma}{\pi\bar n}e^{-\frac{|\gamma|^2}{\bar n}}\sum_{m,n}\sqrt{\frac{\bar n^{m+n}}{(\bar n +1)^{m+n+2}}}\times\nonumber\\
& \times& \langle n|\gamma+\eta_t\rangle\langle\gamma+\eta_t|m\rangle \ket n\bra m \label{A2}
\end{eqnarray}
and the scalar products above read:
$\langle n|\gamma+\eta_t\rangle=\exp \left(- |\gamma+\eta_t|^2/2 \right)\left( (\gamma+\eta_t)^n/\sqrt{n!}\right).$
The strategy is now to use this relation and rewrite each sum in (\ref{A2}) as a coherent state but with a rescaled argument,
and then try to rewrite (\ref{A2}) as a single thermal state (with a different mean photon number than $\varrho_0$).
To this end we note that:
\begin{eqnarray}
&&e^{-\frac{1}{2}|\gamma+\eta_t|^2}\sum_n\left(\frac{\bar n}{\bar n +1}\right)^{\frac{n}{2}}\frac{(\gamma+\eta_t)^n}{\sqrt{n!}}\ket n=\\&&e^{-\frac{1}{2}\frac{|\gamma+\eta_t|^2}{\bar n+1}}\left|\sqrt{\frac{\bar n}{\bar n +1}}(\gamma+\eta_t)\right\rangle.
\end{eqnarray}
Substituting this into (\ref{A2}) and reordering gives:
\begin{eqnarray}
&&\hat A_t 
=\frac{1}{\bar n +1}e^{-\frac{|\eta_t|^2}{1+2\bar n}}\int\frac{d^2\gamma}{\pi\bar n}e^{-\frac{1+2\bar n}{\bar n(\bar n +1)}
\left|\gamma+\frac{\bar n}{1+2\bar n}\eta_t\right|^2}\times\nonumber\\
&&\times \left|\sqrt{\frac{\bar n}{\bar n +1}}(\gamma+\eta_t)\right\rangle\left\langle\sqrt{\frac{\bar n}{\bar n +1}}(\gamma+\eta_t)\right|.\label{A3}
\end{eqnarray}
Note that since we are interested in $tr\sqrt{\hat A_t}$ rather than $\hat A_t$ itself, there is a freedom 
of rotating $\hat A_t$ by a unitary operator, in particular by a displacement.
We now find such a displacement as to turn (\ref{A3}) into the thermal form.
Comparing the exponential under the integral in (\ref{A3}) with the thermal form, we see that the argument of the subsequent coherent states
should be proportional to $\gamma+\left(\bar n\right)/\left(1+2\bar n\right) \eta_t$. Simple algebra gives:
\begin{eqnarray}
&&\left|\sqrt{\frac{\bar n}{\bar n +1}}(\gamma+\eta_t)\right\rangle\simeq\\
&&\hat D\left(\sqrt{\frac{\bar n}{\bar n +1}}\frac{\bar n+1}{1+2\bar n}\eta_t\right)\left|\sqrt{\frac{\bar n}{\bar n +1}}\left(\gamma+\frac{\bar n}{1+2\bar n}\eta_t\right)\right\rangle\nonumber,
\end{eqnarray}
where we have omitted the irrelevant phase factor  as we are interested in the projector on the above state.
Inserting the above relation into (\ref{A3}), dropping the displacements, and changing the integration variable:
$\gamma \to \sqrt{\bar n/\left(\bar n +1\right)}\left(\gamma+\left(1+2\bar n\right)\eta_t\right)$
gives:
\be
B^{mic}_{X,X'}(t)=e^{-\frac{1}{2}\frac{|\eta_t|^2}{1+2\bar n}}\frac{1}{\sqrt{1+2\bar n}}tr\sqrt{\varrho_{th}\left(\bar n^2/(1+2\bar n)\right)},
\ee
where $\varrho_{th}(\bar n)$ is a thermal state with the mean photon number $\bar n$. We 
 use the Fock expansion for $\varrho_{th}\left(\bar n^2/(1+2\bar n)\right)$:
\begin{eqnarray}
&&B^{mic}_{X,X'}(t)=e^{-\frac{|\eta_t|^2}{2+4\bar n}}\frac{1}{\sqrt{1+2\bar n}}\times\\
&&\times\left(1+\frac{\bar n^2}{1+2\bar n}\right)^{-\frac{1}{2}}
\sum_n \left(\frac{\bar n^2/(1+2\bar n)}{1+\bar n^2/(1+2\bar n)}\right)^{\frac{n}{2}}\\
&&=\exp\left[-\frac{(X-X')^2}{2}|\alpha(t)|^2\text{th}\left(\frac{\beta\omega}{2}\right)\right],
\end{eqnarray}
where we have used the definition of $\eta_t$ and $\bar n=1/(e^{\beta\omega}-1)$. 
Coming back to the generalized overlap for macro-fraction states (\ref{Bm}) with a help of (\ref{B}),
we finally obtain the desired result (\ref{BT}).

\acknowledgments
We would like to thank R. Horodecki, P. Horodecki, J. Wehr, H. Gomonay, M. Lewenstein, P. Massignan and A. Lampo 
for discussions.
JT was supported by the ERC Advanced Grant QOLAPS, JKK and JT acknowledge the financial support of   
National Science Centre project Maestro DEC-2011/02/A/ST2/00305.

\end{document}